\documentclass[12pt]{article}

\usepackage{a4}
\usepackage{a4wide}

\usepackage{amsmath}
\usepackage{amscd}
\usepackage{amsfonts}
\usepackage{amssymb}
\usepackage{mathrsfs}
\usepackage{fancybox} 
\usepackage{enumerate}
\usepackage{accents} 

\usepackage{color}
\usepackage{ulem}

\def\half{{1\over2}}

\def\={\stackrel{\bullet}{=}}

\def\({\left(}
\def\){\right)}
\def\[{\left[}
\def\]{\right]}

\def \be {\begin{equation}}
\def \ee {\end{equation}}
\def \beqa {\begin{eqnarray}}
\def \eeqa {\end{eqnarray}}
\def \beal#1 {\begin{align}#1\end{align}}
\def \bes#1 {\begin{equation}\begin{split}#1\end{split}\end{equation}}
\def \nn {\notag\\}

\begin{document}

\begin{titlepage}
\title{
\vspace{-2cm}
\begin{flushright}
\normalsize{ 
YITP-20-59 \\ 
OU-HET-1056
}
\end{flushright}
       \vspace{1.5cm}
Conserved charges in general relativity
       \vspace{1.cm}
}
\author{
 Sinya Aoki\thanks{saoki[at]yukawa.kyoto-u.ac.jp },
\; Tetsuya Onogi\thanks{onogi[at]phy.sci.osaka-u.ac.jp},\; 
Shuichi Yokoyama\thanks{shuichi.yokoyama[at]yukawa.kyoto-u.ac.jp},\; 
\\[25pt] 
${}^{*}{}^{\dagger}{}^{\ddagger}$ {\normalsize\it Center for Gravitational Physics,} \\
{\normalsize\it Yukawa Institute for Theoretical Physics, Kyoto University,}\\
{\normalsize\it Kitashirakawa Oiwake-cho, Sakyo-Ku, Kyoto 606-8502, Japan}
\\[10pt]
${}^\dagger$ {\normalsize\it Department of Physics, Osaka University,}\\ 
{\normalsize\it Toyonaka, Osaka 560-0043, Japan,}
}

\date{}

\maketitle

\thispagestyle{empty}


\begin{abstract}
\vspace{0.3cm}
\normalsize

We present a precise definition of a conserved quantity from an arbitrary covariantly conserved current available in a general curved spacetime 
with Killing vectors.
This definition enables us to define energy and momentum for matter by the 
volume integral.
As a result we can compute charges of Schwarzschild and BTZ black holes by the volume integration of a delta function singularity. 
Employing the definition we also compute the total energy of a static compact star. It contains both the gravitational mass known as the Misner-Sharp mass in the Oppenheimer-Volkoff equation and the gravitational binding energy. We show that the gravitational binding energy has the negative contribution at maximum by 68\% of the gravitational mass in the case of a constant density.
We finally comment on a definition of generators associated with a vector field on a general curved manifold.

\end{abstract}
\end{titlepage}

\section{Introduction} 
\label{Intro} 

Since Einstein submitted papers on general relativity \cite{Einstein:1916}, classical or quantum field theory on a curved spacetime has extensively been investigated. When spacetime is curved, the physical quantities defined on flat spacetime are required to be modified suitably in accordance with general covariance. 
For example, a conserved current, which exists in the presence of global symmetry in the system \cite{Nother:1918}, is modified to be a covariantly conserved one on a general curved spacetime. 

However there has been no general argument to define a conserved charge from a covariantly conserved current, which inevitably causes a problem to define energy and momentum. 
Einstein originally argued that the conservation law of energy and momentum for matter follows as long as they are combined with those for gravitational field \cite{Einstein:1916}. (See also \cite{Landau:1975}.)
The corresponding energy momentum tensor for the gravitational field, however, is not covariant under general coordinate transformation. As a result an energy defined as in the case of flat spacetime depends on a coordinate system and conserves only in the particular frame. 

One way to circumvent this issue is to define an energy locally on the asymptotic region of spacetime called quasi-local energy. 
Initially the quasi-local energy and momentum were studied on an asymptotically flat spacetime by recasting gravity system into the Hamiltonian dynamics known as the ADM formalism \cite{Arnowitt:1962hi}. (See also \cite{Bondi:1962px}.) They are defined by a surface integral in the asymptotic region, by which the invariance under a class of general coordinate transformations preserving a boundary condition was achieved. 
This result has been further extended for a more general curved spacetime with surface terms suitably incorporated 
\cite{Brown:1992br,Hawking:1995fd,Horowitz:1998ha,Balasubramanian:1999re,Ashtekar:1999jx}. 
A caveat in this extension is that boundary terms accompany with divergence even in the flat spacetime, so that one needs to subtract it by comparing a reference frame or by adding local counter terms.

The authors of the present paper investigated a property of a black hole holographically realized by the flow equation method \cite{Aoki:2020ztd}. 
In the study we encountered a situation to evaluate the energy of the total system with matter spread all over the space, which is required to be evaluated by the volume integral of the energy density.
We  reached a manifestly covariant  definition of a conserved charge from a covariantly conserved current  in a general curved spacetime, which 
extends the one in Ref.~\cite{Fock1959},  and
improves the one given in \cite{Komar:1962xha,Abbott:1981ff} for special backgrounds. 
This allows us to define energy and momentum for matter in a form of the volume integral at an arbitrary time slice of a given curved spacetime.
A virtue is to enable one to evaluate charges of black holes just like an electric charge of the electron in electro-magnetism by the volume integral of 
an integrable singularity such as the delta function.
Furthermore, applying
our definition to the 
energy
of a compact star,
we
discover a correction to the mass formula obtained from the Oppenheimer-Volkoff equation, 
which represents a contribution from the gravitational interaction and becomes
68\% of the mass 
at the maximum
for a constant density.

\section{ Conserved charge from covariantly conserved current}
\label{sec:Proposal}

Consider any classical or quantum field theory on a general curved spacetime.  
Suppose there exists a covariantly conserved current $J^\mu$, $\nabla_\mu J^\mu = 0$, where $\nabla_\mu$ is the covariant derivative for the metric $g_{\mu\nu}$. 
Then we claim that the following quantity is conserved under the given time evolution
\be
Q(t) := \int_{M_t} d^{d-1}\vec x\,  \sqrt{\vert g\vert} J^0 (t,\vec x), 
\label{eq:def_q}
\ee
where $M_t$ represents a time slice of the spacetime $M$ at the time $t$, $g$ denotes the determinant of $g_{\mu\nu}$, and $d$ is the dimension of the spacetime $M$. If there exists boundary for $M_t$, we set the boundary condition for
the fields to fall off sufficiently fast at boundary of $M_t$ for all $t$. We emphasize that $g$ is the determinant of the metric in the total spacetime, which contains the time components. 

To show this, we assume the spacetime has the foliation structure for simplicity. 
(The same argument is used in literature. For example, see \cite{Townsend:1997ku}.)
Let us consider the same quantity defined by \eqref{eq:def_q} at another time slice with $t'$ greater than $t$, and take a submanifold $M'$ with the foliation structure whose boundary contains $M_t$ and $M_{t'}$. Such a manifold may be written formally as $M' = \prod_{t\leq s \leq t'} M_{s}$.
Under the boundary condition, the difference between $Q(t^\prime)$ and $Q(t)$ becomes
\beal{ 
Q(t') - Q(t) 
=& \int_{M'} d^{d} x\, \partial_\mu( \sqrt{\vert g\vert} J^\mu (t,\vec x) ) = 0,
}
where we used
$\partial_\mu( \sqrt{\vert g\vert} J^\mu (t,\vec x) ) =\sqrt{\vert g\vert} \nabla_\mu J^\mu (t,\vec x) =0$. 
This proves that $Q(t)$ is independent of $t$. 
The charge $Q$ is a scalar under the assumption, though the generalization is straightforward. Note  that $Q$ is not a scalar if it is defined from a higher rank tensor rather than the vector.

This formula can be applied to the computation of a conserved charge for any gravitational systems with a Killing vector. A covariantly conserved current associated with a Killing vector $\xi$ can be constructed as
\be 
J^\mu = T^\mu{}_\nu \xi^\nu, 
\label{currentKilling} 
\ee
where $T^\mu{}_\nu$ is the given energy momentum tensor for matter. It can be easily shown that this is covariantly conserved by using $\nabla_\mu T^\mu{}_\nu =0$ and $\nabla_{\mu} \xi_{\nu}+\nabla_{\nu} \xi_{\mu} = 0$. 
Note that the definition of a conserved  charge using eq.~\eqref{currentKilling} in $d=4$ appears in \cite{Fock1959,Trautman:2002zz}, but the formula has been rarely used in literature as far as we know.
If $\xi^\mu$ is a Killing vector associated with the time translation, the conserved charge 
becomes the total energy of the system
\beqa
E = \int_{M_t} d^{d-1}\vec x \sqrt{\vert g\vert} T^0{}_\mu \xi^\mu,
\eeqa
which 
agrees with the standard definition of the energy
in the flat 
background with $\xi^\mu=-\delta^\mu_0$.
In the next section we compute charges of several black holes by using this formula.

\section{Conserved charges of black holes} 
\label{sec:Blackhole}

In this section we compute a conserved charge for Schwarzschild and BTZ black holes employing the presented formula. 

\subsection{ Schwarzschild black hole}

In order to explain the key idea of the calculation we start with the simplest setup. That is, 
we begin with the Einstein equation 
\be 
R_{\mu\nu} - \frac{1}{2} g_{\mu\nu} R + \Lambda g_{\mu\nu} = 0. 
\label{EOMV}
\ee
This is satisfied by the Schwarzschild black hole solution: 
\beal{
d s^2 =& -f(r) (dx^0)^2  + {1\over f(r) } dr^2  + r^2\tilde g_{ij} dx^i dx^j,
\label{eq:SBH}
}
where $r$ is the radial coordinate and the $d-2$ dimensional manifold fibered over the cone is an Einstein manifold, whose Ricci tensor is given by ${}^{(d-2)} R_{ij} = (d-3) k \tilde g_{ij}$ with a constant $k$, and 
\be
f(r) =\frac{-2 \Lambda r^2}{(d-2)(d-1)} + k - \frac{2G_N M }{r^{d-3}}. 
\label{f}
\ee
Note that for a positive or non-positive $k$ the submanifold is compact or non-compact, respectively.

Since this is a static solution, there exists a Killing vector with $\xi^\mu=-\delta^\mu_0$, which corresponds to the time translation. 
Thus the corresponding charge is the energy of the system:
\be 
E= \int d^{d-1}\vec x\,  \sqrt{\vert g\vert} (-T^0{}_0), 
\label{mass}
\ee
where the matter energy momentum tensor is given by
\be 
T_{\mu\nu}
= {1\over 8\pi G_N} ( 
R_{\mu\nu} - \frac{1}{2} g_{\mu\nu} R + \Lambda g_{\mu\nu} ).
\ee
with $G_N$ the Newton constant.
According to the equation of motion \eqref{EOMV} this energy momentum tensor seems to vanish on shell, but it does not. We emphasize that it vanishes except a singularity located at $r=0$. 
This singularity contributes to the charge.

In order to evaluate the contribution, 
we first compute the energy by expanding the energy momentum tensor in terms of the weak field around $r=\infty$. That is,
we write 
$ f=\bar f +\delta f$ with $\delta f = -2G_N M/r^{d-3}$
and expand the stress tensor perturbatively around $\bar f$ 
to extract a pole.
This can be done by separating the metric into the regular part $\bar g_{\mu\nu}$ and the singular part $h_{\mu\nu}$, the latter of which is given by
\beqa
h_{\mu\nu} dx^\mu dx^\nu &=& -\delta f (dx^0)^2 +\left(\frac{1}{f} -\frac{1}{\bar f}\right) dr^2.
\eeqa  
At the leading order, we  have
\beqa
T^0{}_0
&=&-\half \left( {1\over \sqrt{|\bar g|}} \partial_\mu( \sqrt{|\bar g|} \bar g^{\mu\nu} \partial_\nu h_{0}^0) - \bar f^{-1}\bar f'^2 h^0_{0} \right) 
+  \bar\nabla^{0} \bar\nabla_\sigma h^\sigma_{0}- h_0^0 -\half \bar\nabla_{\mu} \bar\nabla_\sigma h^{\sigma\mu} + \cdots ,~~~~~
\eeqa
where $\bar\nabla_\mu$ is the covariant derivative with respect to the metric $\bar g_{\mu\nu}$,
and the ellipsis represents the higher order terms with respect to $h$. 
This must vanish except at the origin, and indeed this can be written as  
\beqa
T^0{}_0 &=&  \frac{d-2}{16\pi G_N r^{d-2}}\partial_r \left( r^{d-3}\delta f\right) + \cdots , 
\label{eq:T00}
\eeqa
which has the desired property. 
Plugging this into \eqref{mass} we can compute the charge as 
\beqa 
E
&=&- \int d^{d-1}\vec x\,  \sqrt{\vert \tilde g\vert} \frac{d-2}{16\pi G_N}\partial_r \left( r^{d-3} \delta f \right) 
= \rho V_{d-2},~~~
\label{massS}
\eeqa
where $V_{d-2} = \int d^{d-2} x \sqrt{\vert \tilde g\vert }$ is the volume of the Einstein manifold
with $\tilde g$ being the determinant of $\tilde g_{ij}$, and
$\rho = (d-2) M/(8\pi)$ is the energy or mass density. To evaluate $r$ integral we employ the Stokes' theorem.
Note that the higher order terms do not contribute to the surface integral. 
Our result reproduces the known result obtained by other methods. (For example, see eq.(2.5) in Ref.~\cite{Horowitz:1998ha}.)

On the other hand, it is also possible to compute the contribution of the singularity by the direct calculation of the Ricci tensor. 
From the direct calculation one finds that the
first term in Eq.~\eqref{eq:T00} is indeed exact. 
More detailed results are as follows.
\beqa
R^0{}_0 &=& -\frac{1}{2r^{d-2}}\partial_r\left( r^{d-2}\partial_r f(r) \right) = R^r{}_r,\\
R^i{}_j &=& \delta^i_j\left[\frac{(d-3)k}{r^2} -\frac{1}{r^{d-2}}\partial_r\left(r^{d-3} f(r)\right)\right], 
\eeqa
which leads to
a form of $T^\mu{}_\nu$ proportional to a delta function:
\beqa
 T^0{}_0 &=&  \frac{d-2}{16\pi G_N r^{d-2}}\partial_r  \left( r^{d-3} 
 \delta f(r)\theta (r)
 \right)
 = - \rho \frac{\delta(r)}{r^{d-2}} =T^r{}_r, \nn 
 T^i{}_j&=& -\delta^i_j \frac{\rho}{d-2} \frac{r\partial_r\delta(r)}{r^{d-2}}
 = \delta^i_j \frac{\rho}{d-2} \frac{\delta(r)}{r^{d-2}}.
  \label{eq:T00B}
\eeqa
We here inserted the step function $\theta$ with $\theta(0)=0$ for the singular term in \eqref{f} 
to explicitly extract the singular contribution, 
and used $\delta(r)=\frac{d \theta(r)}{d r} $.
Writing $\delta(r)$ in terms of $\delta^{(d)}(\vec x)$, this agrees with a result at $d=4$ by distributional techniques\cite{Balasin:1993fn}.
Thus the matter energy momentum tensor $T^\mu{}_\nu$ for the Schwarzschild black hole can be understood as a distribution.

Although we do not encounter any mathematical problem to derive this result, one may wonder the physical validity to perform the volume integral of the constant $x^0$ slice, which becomes time-like inside the horizon.\footnote{
We would like to thank M.~Sasaki, T.~Shiromizu and S.~Sugimoto for raising this question and for discussion using the Penrose diagram.
}
To clear up this subtlety, let us consider a simplified situation where the cosmological constant vanishes with $k=1$. Then we can move to the Eddington-Finkelstein coordinates \cite{Eddington:1924aa,Finkelstein:1958zz}
\beqa
ds^2 = -(1+u) d\tau^2 -2u d\tau dr +(1-u) dr^2 +r^2\tilde g_{ij} dx^idx^j,
\eeqa
where $u =-2G_N M/r^{d-3}$ and we changed the time variable from $x^0$ to $\tau = x^0 -g(r) $ with  $
\frac{d g(r)}{d r} =\frac{u}{1+u}$. Then the unit normal vector to the constant $\tau$ slice defined by $n_\mu= -{1\over \sqrt{ 1 - u}} \delta^\tau_\mu$ is always well-defined and time-like for any nonzero $r$. 
The conserved energy, which takes the same form due to its manifest general covariance, is computed as
\beqa
E &=& -\int_{M_\tau} d^{d-1}\vec x\, \sqrt{\vert g\vert} T^\tau{}_\tau = \rho V_{d-2},
\eeqa
which agrees with Eq.~\eqref{massS},
where $T^\tau{}_\tau = \frac{(d-2)}{16\pi G_N} \frac{\partial_r( r^{d-3} u)}{r^{d-2}}.$

\subsection{Reissner-Nordstr\"om black hole} 
Below we perform the same computation of a mass of a charged black hole in general $d$ dimensions,
for which the the metric is given in the same form as \eqref{eq:SBH} except that  $f(r)$ is replaced by  $f_q(r) = f (r)+{d-3 \over d-2} 8\pi G_N q^2 r^{-2(d-3)}$, 
together with the gauge potential 
$A_\mu = (-\frac{q}{r^{d-3}} +\frac{q}{r_+^{d-3}}) \delta^0_{\mu}
$, where $q, r_+$ are constants\cite{Chamblin:1999tk}.
This configuration of gravitational and gauge fields satisfies the equations of motion given by
\beqa
G_{\mu\nu} +\Lambda g_{\mu\nu} &=&  
8\pi G_N ( T^G_{\mu\nu} +T^A_{\mu\nu}),
~ \nabla_\mu F^\mu{}_{\nu} =J_\nu, ~~~
\eeqa
where
$F_{\mu\nu} :=\nabla_\mu A_\nu -\nabla_\nu A_\mu $ and
$T^A_{\mu\nu} := F_{\mu}{}^\alpha F_{\nu\alpha} -\frac{1}{4} g_{\mu\nu} F_{\alpha\beta} F^{\alpha\beta}$.
Here $ T^G_{\mu\nu}$ and $J_\nu $ explicitly represent the singular contributions of the metric and the gauge potential at $r=0$, respectively.
Explicitly $(T^G + T^A)^0{}_0$ is given in \eqref{eq:T00B} by replacing $\delta f$ with
$\delta f_q= \delta f + {d-3 \over d-2} 8\pi G_N q^2 r^{-2(d-3)}$.
This also agrees with the distributional result at $d=4$\cite{Balasin:1993kf}.

Since this metric is also static, the energy defined by \eqref{mass} is conserved. 
However this charge diverges, due to the contribution of the electromagnetic field.
Physically, this divergence can be interpreted as 
that of
the self-energy for the charged point particle.
Indeed it remains even for the flat space-time with $M=0$ and $\Lambda=0$.
Therefore,
classically,  the charged black hole has the infinite energy due to the infinite electromagnetic
energy. Thus the renormalization as well as the quantization of the gauge field on the curved space are needed to fix this problem, as is the case
 on the flat space.

Fortunately, since $\nabla_\mu(T^G)^{\mu}{}_ 0 =0$ (thus  $\nabla_\mu(T^A)^{\mu}{}_ 0 =0$),  we can define an energy from the covariantly conserved $T^G$  alone without electromagnetic energy as
\beqa
(T^G)^0{}_0 &=& -\frac{(d-2)}{16\pi G_d r^{d-2} }\partial_r\left(r^{d-3} \delta f(r)\right),
\eeqa
where  $\delta f$ is given before.
We thus obtain
\beqa
E_G &=& \int d^{d-2} \vec x\,\int dr \sqrt{\vert g\vert} (T^{G})^0{}_0 \xi^0 = V_{d-2}\rho ,
\eeqa
which reproduces the result in 
\cite{Chamblin:1999tk} 
as a special case choosing a sphere as the internal manifold.

This system allows another conserved quantity, thanks to the invariance  under the $U(1)$ gauge transformation by $\delta A_\mu =\partial_\mu \theta$, which leads to
\beqa
\partial_\mu j^\mu &=& 0, \quad j^\mu = \nabla_\nu \left(\sqrt{\vert g\vert} F^{\mu\nu}\right)
\eeqa
without using the Maxwell equation.  
According to our prescription, $Q_c= \int d^{d-2} x\,\int dr\, \sqrt{\vert g\vert} J^0$
with $J^0=j^0/\sqrt{\vert g\vert}$
gives
the conserved electric charge, which is evaluated as $Q_c = V_{d-2} (d-3) q .$
At $d=4$ for $k>0$, $Q_c= 4\pi q$.

\subsection{BTZ black hole} 

As a final example, we compute a charge different from a mass. 
To this end we consider a BTZ black hole and compute its angular momentum \cite{Banados:1992wn}. 
The metric 
\be 
ds^2 = -f(r) dt^2 + {1\over f(r)} dr^2 + r^2(d\phi -\omega(r)dt) ^2, 
\ee
where 
\be
f(r)= {r^2 \over L^2} - 2G_N M \theta(r) + {G_N^2 J^2 \over 4r^2} , ~~ \omega(r) = {G_N J \over 2 r^2} , 
\ee
with $M,J$ are constants, 
satisfies the Einstein equation in three dimensions. We insert the step function to the constant part to emphasize that this solution is valid except the origin. 

This BTZ black hole has not only a Killing vector with respect to the time translation but also the one which rotates the system, $\xi^\mu = \delta^\mu_\phi$. 
As in the previous cases the first one defines the mass, which can be similarly computed as $E=\frac{M}{4}$. 
On the other hand, the second Killing vector define an angular momentum: 
\beal{
P_\phi=& \int d^{2}x \sqrt{|g|} T^0{}_{\phi}. 
}
$T^0{}_{\phi}$ is computed from the Einstein tensor as 
$T^0{}_{\phi} 
=-\frac{1}{ 16\pi G_N r} \partial_r \left(r^3 \omega'(r) \right) $. Thus we find $P_\phi ={J\over 8} $, which reproduces the known result \cite{Banados:1992wn}.

\section{Mass of a compact star}

In this section we apply a conserved charge to the computation of the total energy of a static compact star setting $k=1$ with $d\geq3$. 
\subsection{Oppenheimer-Volkoff equation}
Let us consider 
a spherically symmetric system such that the metric is given by
\beal{
d s^2 =& -f(r) (dx^0)^2  + {h(r) } dr^2  + r^2\tilde g_{ij} dx^i dx^j,
\label{star}
}
and the matter energy momentum tensor is described by the perfect fluid as 
\beqa
T^0{}_0 &=& -  \rho(r), \quad 
T^r{}_r =  P(r), \quad
T^i_j = \delta^i_j P(r),
\label{perfectFulid}
\eeqa
where $\rho(r)$ is the 
density and $P(r)$ is the pressure. 
From the Einstein equation, we can derive useful formulas 
\beal{
P(r)+\rho(r) =& \frac{(d-2) \left(\frac{h'(r)}{h(r)} +\frac{ f'(r)}{f(r)} \right)}{16\pi G_N r h(r)}, ~~~ 
P'(r)   = -\frac{(P(r)+\rho(r))}{2} \frac{f'(r)}{f(r)}. 
\label{useful}
}
From these equations we can derive
the Oppenheimer-Volkoff 
or TOV
equation 
\cite{Oppenheimer:1939ne,Tolman:1939jz}
\beqa
\frac{d P(r)}{d r} &=&- \frac{ G_N \rho(r) M(r)}{ r^{d-2}} 
\Biggl( 1 +\frac{P(r)}{\rho(r)}\Biggr) h(r)
 \left\{ d-3+ \frac{ r^{d-1}}{(d-2)M(r)}
 \Biggl(8\pi P(r)  -\frac{2\Lambda}{(d-1) G_N} \Biggr)\right\},\nn
\eeqa
where 
\beqa
\frac{1}{h(r)} &:=&\frac{-2\Lambda r^2}{(d-2)(d-1)} + 
1 -\frac{2G_N M(r)}{ r^{d-3}} ,
\eeqa
and
\beqa
M(r) &=& \frac{8\pi}{d-2} \int_0^{r} ds\, s^{d-2} \rho(s).
\label{MS}
\eeqa
In order for this system to describe a static compact star we impose a boundary condition such that the pressure vanishes and the energy momentum tensor  is covariantly conserved at the surface of the star.
This makes the pressure and the density vanish outside the star, so that the metric becomes the Schwarzschild outside the star
, namely 
$f(r) =1/h(r)$ with $M=M(R)$ in \eqref{eq:SBH}, where $R$ is the radius of the star.
We here remark that this model does not admit the zero radius limit with positive $M(R)$ fixed, since the stress tensor of the Schwarzschild black hole given by eq.~\eqref{eq:T00B} is not the form of the perfect fluid, \eqref{perfectFulid}.
This shows that there needs to be a certain dynamical process for a star described by \eqref{star} and \eqref{perfectFulid} to collapse into the Schwarzschild blackhole. We shall confirm this explicitly in the analysis for constant density.

\subsection{Total energy with an equation of state
} 
We define the total energy of this system 
by the conserved charge corresponding to the Killing vector $\xi^\mu = -\delta^\mu_0$.
\footnote{
There is a traditional argument for the definition of the total energy of a compact star and its
interpretation, which is different from our result presented below. 
Just for clarity and to avoid confusion, we present our argument in main text and make the comparison to a traditional result in appendix.}
Then it can be computed as 
\beqa
E 
&=& V_{d-2} \int_0^R dr \sqrt{f(r)h(r)}r^{d-2}  \rho(r).
\label{TotalEnergy}
\eeqa
Employing \eqref{useful}, \eqref{MS} ,
we find 
\beal{
E 
=&  \frac{(d-2) V_{d-2}}{8\pi} ( M(R) +\Delta M ), \label{eq:NS_mass}
\\
\Delta M=& -\frac{8\pi G_N}{d-2} \int_0^R dr\, \sqrt{f(r) h(r)^3} r M(r)\{\rho(r) + P(r)\} .
\label{binding}
}
The first leading term gives the gravitational mass of the star also called as the Misner-Sharp mass\cite{Misner:1964je}, which is directly related to an actual observable quantity appearing in the metric around the spacial infinity, 
while the second term is a deviation from the gravitational mass, which corresponds to the gravitational binding energy as we shall see. 

In order to investigate the deviation term in more detail,\footnote{
We would like to thank M.~Shibata, S.~Yamaguchi and S.~Mukohyama to motivate us to study the deviation from the Misner-Sharp mass.
} let us impose the matter consisting of the star to satisfy an equation of state 
\be 
P(r) = w\rho(r).
\label{EoS}
\ee
Plugging this into \eqref{useful} leads to 
$f= \rho^{-\frac{2w}{(1+w)} },$ where we fixed the integration constant to satisfy $f\to1$ in the Newtonian limit $w\to0$.\footnote{
This integration constant can be also absorbed by rescaling the time variable.
} 
Then the deviation term becomes  
\be
\Delta M  = - \frac{8\pi G_N (1+w)}{d-2} \int_0^R dr r  \rho^{\frac{1}{(1+w)} }{1\over \left({-2\Lambda r^2 \over (d-1)(d-2)} + 1 - {2 G_N M(r) \over r^{d-3}}\right)^{3/2} }   M(r) .
\ee
Let us expand this in terms of the cosmological constant and the Newton constant. 
The leading and next-to leading terms are given by 
\beal{
\Delta M_1  =& - \frac{8\pi G_N (1+w)}{d-2} \int_0^R dr r  \rho^{\frac{1}{(1+w)} }   M(r),   \\
\Delta M_2  =& - \frac{24\pi G_N (1+w)}{d-2} \int_0^R dr  \rho^{\frac{1}{(1+w)} } \left({\Lambda r^3 \over (d-1)(d-2)} + { G_N M(r) \over r^{d-4}}\right)  M(r). 
}
Here $\Delta M_1$ represents the leading order of the gravitational self-interaction energy of the matter inside the star obeying the equation of state \eqref{EoS}.
In particular, at four dimensions in the Newtonian limit $w\to 0$, this can be rewritten as 
\beqa
\Delta M_1 =  -\frac{G_N}{2}  \int 
d^3x\, d^3y  \, \frac{\rho(\vec x) \rho(\vec y)}{\vert \vec x-\vec y\vert} ,
\eeqa
which is nothing but Newtonian gravitational energy inside the star including the symmetric factor $1/2$.

\subsection{Estimation with constant density}
\label{constantDensity} 
In order to estimate the size of the correction $\Delta M$,
let us consider a case of constant density $\rho(r) = \rho_0$ and vanishing cosmological constant. 
In this case, we can compute $M(r)$ given in \eqref{MS} as $
M(r) = \frac{8\pi r^{d-1}\rho_0}{(d-2)(d-1)}$, which leads to  $h(r)^{-1} = 1 - {r^2 \over r_0^2}$ with $r_0 =\sqrt \frac{(d-2) (d-1)}{16\pi G_N \rho_0 } $. 
The TOV equation with the boundary condition is easily solved as 
\beal{
P(r)=&\frac{\rho_{0} (d-3) \left(\sqrt{r_0^2 - r^2}-\sqrt{r_0^2 - R^2}\right) }{-(d-3) \sqrt{r_0^2 - r^2} +(d-1)   \sqrt{r_0^2 - R^2}}. 
}
For a stable star the pressure has to be finite for all $r$. 
 This leads to an inequality\footnote{
This inequality leads to the lower bound for the radius of the star $R_{\rm min}$ as $
R^{d-3} \ge  \frac{(d-1)^2}{2(d-2)} G_N M_0 =R_{\rm min}^{d-3}$ for 
keeping $M(R)=M_0$ independent on $R$. 
This is consistent with the previous argument that this system does not admit the zero size limit fixing a positive $M(R)=M_0$.
} 
\be 
R < \frac{2 \sqrt{d-2} }{d-1 }r_0 =:R_{*}.
\ee
$\Delta M$ can be written as 
\beal{
\Delta M
=-{(d-1) \sqrt{r_0^2 -R^2} \over 4G_N r_0^2 } \int_0^R dr { r^d  \over (r_0^2 -r^2)^{3/2}}.
}
At four dimensions, this is computed as 
\be 
F(R):= \frac{\Delta M}{M(R)}
=-\frac{3}{8G_NM(R)} \left(-\frac{R^3}{r_0^2}+3 R-3 \sqrt{r_0^2-R^2} \arctan\left(\frac{R}{\sqrt{r_0^2-R^2}}\right)\right),
\ee
where $F(R)$ turns out to be a monotonically decreasing function 
from $F(0)=0$ to $F(R_{*})\simeq - 0.68$. 
Therefore the total energy $E$ can be about one third of $M(R)$ so that the gravitational binding energy could be considerably large. 
In particular the total energy of a static compact star becomes smaller than the gravitational mass observed at spacial infinity due to the gravitational self-interaction.

\section{Discussion}
\label{sec:Discussion}
We have proposed a general definition of a conserved charge from any covariantly conserved current,
which requires no specific asymptotic behaviors/approximations for the metric, or no subtraction of boundary contributions, 
as long as a Killing vector exists.
Our definition has reproduced the mass, electric charge and angular momentum of black holes known in the literature.  
Since the presented
formula requires the
matter energy momentum tensor to define the mass,
it is clear that black holes inevitably
have non-zero matter
energy momentum tensor at singularity.
(See also \cite{Buoninfante:2018rlq} in the case of higher derivative gravity.)
One of important consequences in this paper is that any black hole is not a vacuum solution in this treatment.
\footnote{ Traditionally black holes including the ones dealt with in this paper may have been regarded as vacuum solutions in spacetime with their singularity deficient. The method presented in this paper does not return inconsistent results with the traditional treatment, but is applicable to any gravitational object on a general curved spacetime such as a black hole and a compact star.}
This is similar to an electron in electrodynamics: its electric charge distribution is described by a delta function at its position and it is not regarded as a vacuum solution.
We have also demonstrated that the total energy 
of any spherically symmetric compact star defined as the conserved charge cannot be written as a surface term alone, and it contains the gravitational binding energy in addition to the gravitational mass observed at the asymptotic spatial infinity. 

Our definition of charges is formulated as a  generalization of ones used in the flat spacetime to be available in a general curved spacetime. This is achieved to enjoy general covariance manifestly. Therefore our definition has clear physical meaning and is of a generic use with precision compared to other definitions. We discuss difference of a couple of other approaches more in appendix. 

In this paper we focused on a few well-known blackholes such as the Schwarzschild and BTZ ones to compute their charges. In these simple cases the matter stress tensor is
described by a delta function singularity at the origin. It would be interesting to study a more intricate blackhole whose singularity is not point-like any more. In such a case we expect that matter energy momentum tensor is still described by a certain integrable distribution and our definition of charges is valid and useful also as numerical evaluation. 
(Distributional expressions for the Kerr black hole can be found in \cite{Balasin:1993kf}.)

Our proposal is quite generic, so we expect plenty of applications in future. 
As such a potential application, we consider a more  general case where any Killing vectors do not exist. 
We can still consider a charge or a generator associated with a general vector field $\xi^\mu$ as
\be
Q[\xi](t) = \int_{M_t} d^{d-1} \vec x\,  \sqrt{\vert g\vert} T^0{}_\nu \xi^\nu .
\label{chargeglobal}
\ee
Using the similar argument before, we obtain
\beqa
\frac{d Q[\xi]}{d t} &=& \int_{M_t} d^{d-1}\vec x  \sqrt{\vert g\vert}
\rho(x),\label{eq:conv} \quad
\rho (x) := \frac{\left( G^{\mu\nu} +\Lambda g^{\mu\nu}\right) }{16\pi G_N}  (\nabla_{\mu} \xi_{\nu}+\nabla_{\nu} \xi_{\mu})
\eeqa
where  $\rho=0 $ if  $\xi$ is a Killing vector. 
A change of the charge $Q[\xi]$ can be calculated by the volume integral of $\rho$, 
expressed in terms of the gravitational field  through the Einstein equation {with the manifest covariance being kept. 
Eq. \eqref{eq:conv} may give a hint for a {\it  generic conservation equation} 
in general relativity. Note that this does not require any pseudo-tensor. This argument will hold not only for a Lorentzian manifold but also for a more general one.
We will return to this interesting problem in future studies. 

\section*{Acknowledgement}
We would like to thank S.~Mukoyama,  J.~H.~Park,  M.~Sasaki,  M.~Shibata, T.~Shiromizu, S.~Sugimoto and S.~Yamaguchi for useful discussions and 
valuable
comments.
This work is supported in part by the Grant-in-Aid of the Japanese Ministry of Education, Sciences and Technology, Sports and Culture (MEXT) for Scientific Research (Nos.~JP16H03978,  JP18K03620, JP18H05236, JP19K03847). 
S. A. is also supported in part  
by a priority issue (Elucidation of the fundamental laws and evolution of the universe) to be tackled by using Post ``K" Computer, and by Joint Institute for Computational Fundamental Science (JICFuS).
T. O. would like to thank YITP for their kind hospitality during his
stay for the sabbatical leave from his home institute.

\appendix
\section{
Comparison with other definitions
} 
\subsection{The Komar integral}

In this appendix, we compare our results with those obtained by the Komar integral\cite{Komar:1962xha}, defined as
\beqa
E_{\rm Komar}(\xi) =\frac{c}{16\pi G_N} \int_{M_t} d^{d-1}\vec x\, \sqrt{\vert g\vert} \nabla_\mu \nabla^{[0} \xi^{\mu]}
= \frac{c}{16\pi G_N} \int_{\partial M_t} [d^{d-2}\vec x]_\mu\, \sqrt{\vert g\vert}  \nabla^{[0} \xi^{\mu]},
\label{eq:Komar}
\eeqa  
where  $c$ is some constant, $\partial M_t$ is the boundary of
$M_t$ and $ [d^{d-2}\vec x]_\mu$ is its (hyper)surface element normal to the $\mu$ direction. 
The second expression corresponds to the quasi-local definition.

While our result \eqref{massS} for the energy of the Schwarzschild black hole is independent of $\Lambda$, the Komar energy \eqref{eq:Komar} with the Killing vector diverges for 
$\Lambda\not=0$.
For the vanishing cosmological constant, we obtain
\beqa
E_{\rm Komar} (\xi) = \frac{c V_{d-2} (d-3) M}{8\pi }, 
\eeqa
which agrees with Eq.~\eqref{massS} at $d > 3$ if we take $c=(d-2)/(d-3)$.

The divergence in the energy of the charged black hole
also appears in the Komar energy if it is evaluated by the volume integral in the first expression of Eq.~\eqref{eq:Komar}, while such a divergent term vanishes  in the Komar energy evaluated by the surface integral in the second expression of Eq.~\eqref{eq:Komar}.
This explicitly demonstrates that the Stokes' theorem does not hold if the volume integral has divergence, which can not be detected by the surface integral, warning 
that some care is needed to make a conclusion for conserved quantities
by the quasi-local expression. 

The Komar energy for the BTZ black hole diverges due to the non-zero cosmological constant, while the Komar angular momentum agrees with our result, $P_\phi =\frac{J}{8}$, with $c=1$ for a constant.

In the case of a compact star energy, 
the Komar energy at $\Lambda=0$  (otherwise it diverges) 
becomes a pure surface term contribution as
\beqa
E_{\rm Komar} = \frac{c(d-3)V_{d-2}}{8\pi} M(R) .
\eeqa

Let us briefly mention some other evaluations.
At $\Lambda =0$, Ref.~\cite{Brown:1992br} gives
the mass of both neutral and charged black holes as
$E_{\rm BY} =E_{\rm ours}$ at $d=4$ while
the mass of the compact star as  $E_{\rm BY}(R) = M(R)$ at $d=4$.
In the case of the AdS space with $\Lambda < 0$,
Ref.~\cite{Balasubramanian:1999re} gives
\beqa
E=\frac{M}{4}, \quad P_\phi=\frac{J}{8}
\eeqa
for the BTZ blackhole,  $E_{\rm AdS_4}=E_{\rm ours}$ at $d=4$ and 
$E_{\rm AdS_5}=E_{\rm ours} + \frac{3\pi\ell^2}{32 G_N}$ at $d=5$, where $\ell$ is the radius of AdS$_5$ related to the cosmological constant as $\Lambda = -\frac{6}{\ell^2}$. 
The deviation from our result appears after the divergence due to the cosmological constant term is canceled by adding possible quasi-local counter terms.
We also mention that Ref.~\cite{Ashtekar:1999jx} gives the same result as ours for the neutral black hole based on the surface integral using the  'effective stress-energy tensor'.

We summarize the above comparisons in Table~\ref{tab:comp}.

\begin{table}[hbt]
\centering
\begin{tabular}{|l||c|c|c|c|}
\hline
 & Neutral BH &Charged BH & BTZ BH &compact star \\
 \hline\hline
 Ours& $E_{\rm ours}$ & $ E_{\rm ours}+\infty$ &  $E=\frac{M}{4}, P_\phi=\frac{J}{8}$ &
 $E_{\rm ours}(R)+\Delta M$ \\
 \hline
 Komar (quasi-local)  & $E_{\rm Komar}$ & $E_{\rm Komar}+\infty \, (E_{\rm Komar})$ & $E=\infty, P_\phi=\frac{cJ}{8}$ & 
 $E_{\rm Komar}(R)$ \\
\hline
 \cite{Brown:1992br} ($\Lambda=0$, $d=4$)  & $E_{\rm BY}$ & $E_{\rm BY}+ 0$ & -- & $E_{BY}(R)$
 $ $ \\
 \hline
\cite{Balasubramanian:1999re}  ($\Lambda <0, q=0$) &
 \multicolumn{2}{|l|}{$E_{\rm AdS_4} =E_{\rm ours}$,  $E_{\rm AdS_5} =E_{\rm ours}+\delta E$} & $E=\frac{M}{4}, P_\phi=\frac{J}{8}$ & --\\
\hline 
\cite{Ashtekar:1999jx} ($\Lambda <0, q=0$)
& \multicolumn{2}{|l|}
{$E_{\rm AdS_4} = E_{\rm ours}$, $E_{\rm AdS_5} = E_{\rm ours}$ }
&
--
&
--
\\
\hline
\end{tabular}
\caption{A summary of comparisons, where $E_{\rm ours} = \frac{(d-2)V_{d-2}M}{8\pi}$, 
$E_{\rm ours}(R)= \frac{(d-2)V_{d-2}M(R)}{8\pi}$, $E_{\rm Komar} = \frac{c(d-3)V_{d-2}M}{8\pi}$, 
$E_{\rm Komar}(R)= \frac{c(d-3)V_{d-2}M(R)}{8\pi}$, 
$E_{\rm BY} = M$, $E_{BY}(R) =M(R)$, 
and $\delta E =\frac{3\pi\ell^2}{32 G_N}$.
In the case of the Komar energy, we set $\Lambda=0$ except for the BTZ  black hole, otherwise it diverges. The Komar energy for the charged blacked hole is evaulated by volume intergal and by surface integral, where the latter is shown in the parenthesis.
}
\label{tab:comp}
\end{table}

\subsection{The Misner-Sharp mass} 

In this appendix we compare our result to other ones with others} on the total energy of a static compact star at four dimensions and zero cosmological constant. 

Our final result is given by \eqref{TotalEnergy}. (This expression is also written in appendix in \cite{Angus:2018mep}.) 
On the other hand, a traditional argument written in standard  textbooks\cite{Weinberg:1972kfs,Misner:1974qy,Hawking:1973uf,Wald:1984rg,Schutz:1985jx} is that the total energy of a static compact star is given only by the gravitational mass  known as the Misner-Sharp mass\cite{Misner:1964je}, denoted by $M(R)$ in this paper, and that the gravitational binding energy is included in the gravitational mass even though $M(R)$ does not have any non-trivial dependence on the metric as seen from its expression \eqref{MS}. 

According to the textbooks, to compute the gravitational potential energy, one needs to subtract the sum of the static mass and the internal energy from the Misner-Sharp mass: 
\be 
\Omega = M(R) - \varepsilon, 
\label{Omega}
\ee
where $\varepsilon$ is the sum of the static mass and the internal energy. However two different ways to compute $\varepsilon$ are described in the textbooks. One way is to use the volume form in the Cauchy surface \cite{Misner:1974qy,Hawking:1973uf,Wald:1984rg}, which can be written in our convention as 
\beal{
\varepsilon^{(1)} =& \int_0^R \rho \sqrt {\tilde g} d^3x. 
}
The other way is to use the volume form in the total spacetime \cite{Weinberg:1972kfs,Schutz:1985jx}:
\beal{
\varepsilon^{(2)} =& \int_0^R \rho\sqrt {|g|} d^3x. 
}
Note $\varepsilon^{(2)} =E$. (See eq.~\eqref{TotalEnergy}).
What is unsatisfactory in the first method is that it does not respect the general covariance any more. An unsatisfactory point in the second method is that the resulting gravitational potential energy, $\Omega^{(2)}= M(R) - \varepsilon^{(2)}$, becomes positive when the density is constant, because $\Omega^{(2)} = -\Delta M$, where $\Delta M$ is negative for a constant density as shown in section \ref{constantDensity}. For a generic non-constant density, 
$\Omega^{(2)}$ is positive while $\Omega^{(1)}$ is negative in the Newtonian limit\cite{Misner:1974qy,Hawking:1973uf,Wald:1984rg}. 
In addition what is unclear in common to both methods
is to argue that the Misner-Sharp mass, which is independent of the non-trivial metric, contains gravitational binding energy. 
The result presented in this paper has no such an unreasonable point. 

It would be important to scrutinize these results including us from a different perspective such as the tidal force, or the non-removable gravitational force of any extended object.


\end{document}